\documentclass[prd,nofootinbib]{revtex4} 
\usepackage{graphicx} 
\usepackage{amsmath}
\usepackage{amsfonts,amsbsy}
\usepackage{amssymb}

\def\be{\begin{equation}}
\def\ee{\end{equation}}
\def\bea{\begin{eqnarray}}
\def\eea{\end{eqnarray}}

\def\Tr{{\rm tr}}

\begin{document}

\title{\bf On perturbative limits of quadrupole evolution in QCD at high energy}


\author{Jamal Jalilian-Marian$^{1,2}$}
\affiliation{
$^1$ Department of Natural Sciences, Baruch College, CUNY\\
17 Lexington Avenue, New York, NY 10010, USA\\
$^2$ The Graduate School and University Center, City
  University of New York, 365 Fifth Avenue, New York, NY 10016, USA 
}

\begin{abstract}

\noindent We consider the perturbative (weak field) limit of the small $x$ QCD evolution 
equation for quadrupole, the normalized trace of four Wilson lines in the fundamental
representation, which appears in di-hadron angular correlation in high energy collisions. 
We linearize the quadrupole evolution equation and then expand the Wilson lines in 
powers of $g\, A_{\mu}$ where $A_{\mu}$ is the gauge field. The quadratic terms  
in the expansion ($\sim g^2\, A^2$) satisfy the BFKL equation as has been recently 
shown. We then consider the quartic terms ($\sim g^4\, A^4$) in the expansion and 
show that the linearized quadrupole evolution equation, written in terms of color charge density $\rho$, 
reduces to the well-known BJKP equation for the imaginary part of four-reggeized gluon exchange amplitude. 
We comment on the possibility that the BJKP equation for the evolution of a $n$-reggeized gluon state 
can be obtained from the JIMWLK evolution equation for the normalized trace of $n$ fundamental
Wilson lines when non-linear (recombination) terms are neglected.
 
\end{abstract}

\maketitle

\section{Introduction}
The recent experimental observation of disappearance of the away side peak in di-hadron
angular correlation in the forward rapidity region in deuteron-gold collisions at 
RHIC~\cite{star} has generated a lot of interest in multi-parton correlations at high energy (small $x$).
Unlike structure functions in DIS and single inclusive particle production in hadronic
collisions which are sensitive to dipoles (normalized trace of two Wilson lines), di-hadron
correlations probe correlators of higher number of Wilson lines~\cite{jjmyk1,rev}. Therefore one has the
opportunity, for the first time, to investigate these higher correlators experimentally
through studies of angular and rapidity correlations in di-hadron production cross section
in high energy hadronic collisions. Such studies can teach us much about the intrinsic 
correlations in the hadronic or nuclear wave functions which are not accessible in single
inclusive particle production or in studies of structure function in DIS.

Higher correlators of Wilson lines appear in two-hadron production cross section in any
dilute-dense collision at high energy where analytic calculations are possible. Classic 
examples of such asymmetric collisions are proton-nucleus collisions (see~\cite{jjmyk} for a review)  
in the fragmentation region of the proton, and in DIS close to the virtual photon remnants~\footnote{Particle 
production in the very forward rapidity region in proton-proton collisions at very high 
energy falls into this category also.}. Two-gluon production cross section in DIS has been
considered in~\cite{jjmyk1} while two-parton production cross section in proton-nucleus 
collisions has been investigated in~\cite{jjmyk1,akml,corr_cgc,other_qg,sxy}. In all cases, the cross section
involves correlators of higher (more than two) number of Wilson lines, the most important being
the quadrupole operator. Evolution equations for these higher point correlators have been 
derived~\cite{jjmyk1,adjjm,dmmx} and approximate analytic expressions for them have been developed  
using a Gaussian model~\cite{dmxy} and approximate analytic solutions have been proposed~\cite{eidt}. 
Very recently, powerful lattice gauge theory techniques have been applied to solve the JIMWLK evolution 
equation which then allows a systematic and detailed numerical study of the properties of these higher point 
correlators~\cite{djlsv}. 

Here we study the evolution equation for the quadrupole operator in the weak field limit. 
A first study of this has already been performed in~\cite{dmmx} where it is shown that the
quadrupole evolution equation reduces to a sum of BFKL equations for the dipole operator
in the limit where the dipole is expanded in powers of the gluon field and quadratic 
terms in gluon field are kept. Here, we go beyond the quadratic expansion and show that the 
quartic terms in the expansion of the linearized quadrupole evolution equation satisfy an equation which
is identical to the BJKP equation~\cite{multigluon,chen-al} for the imaginary part of the four-reggeized gluon exchange
amplitude. This should be very useful since there is an extensive literature on the properties
of the BJKP equation which may give us further insight on the properties of the JIMWLK equation in the
limit where one may ignore non-linear terms.

\section{Quadrupole Evolution equation}
We start by defining the quadrupole operator $Q$ as 
\be
Q (r, \bar{r},\bar{s},s) \equiv {1\over N_c}  
\Tr V_r\, V^\dagger_{\bar{r}} \, V_{\bar{s}} \, V_s^\dagger 
\label{eq:S_4}
\ee
where $V_r \equiv V (r_t)$ is a Wilson line in the fundamental
representation in the covariant gauge 
\be
V (r_t) \equiv \hat{P} e^{- i g \int d x^- \, A^+}
\label{wilson}
\ee
and $A^\mu (x^-, r_t) = \delta^{\mu\,+} \, \delta (x^-) \alpha (r_t)$. The gauge field $\alpha (r_t)$
is related to the color charge density via $\partial_t^2 \, \alpha^a (r_t) \sim g\, \rho^a (r_t)$
and $r,\bar{r}, \bar{s},s$ etc. denote two-dimensional coordinates on the 
transverse plane. The evolution equation for the quadrupole was derived in~\cite{jjmyk1}
in the large $N_c$ limit and using Feynman diagram techniques. It has been recently 
re-derived~\cite{dmmx} using the JIMWLK equation where it was shown that there are no 
$N_c$ suppressed corrections. Here we outline the derivation using the JIMWLK formalism~\cite{jimwlk} 
where the evolution ($y = log 1/x$) of any operator is given by  

\be
\frac{d}{dy} \langle O \rangle =
  \frac{1}{2} \left< \int d^2x \, d^2y \, \frac{\delta}{\delta\alpha_x^b}
     \, \eta^{bd}_{xy} \, \frac{\delta}{\delta\alpha_y^d} \, O \right>~,
\label{eq:ham}
\ee
with
\be
\eta^{bd}_{xy} = \frac{1}{\pi} \int \frac{d^2z}{(2\pi)^2}
     \frac{(x-z)\cdot(y-z)}{(x-z)^2 (y-z)^2} \left[
       1 + U^\dagger_x U_y - U^\dagger_x U_z - U^\dagger_z U_y
       \right]^{bd} ~.
\label{eq:eta}
\ee
and $U$ is a Wilson line in the adjoint representation. The derivation of the
quadrupole evolution equation is straightforward but tedious. It involves functional 
differentiation of the Wilson lines and repeated use of the identity 
$[U (r)]^{ab} \, t^b = V^\dagger (r) \, t^a\, V (r)$. The result is    
\bea
{d\over dy} \left< Q (r, \bar{r},\bar{s},s)\right> &=& 
{N_c\, \alpha_s \over (2\pi)^2} \int d^2 z \Bigg\{ \Bigg<\!\!  
%
\left[ {(r - \bar{r})^2 \over (r - z)^2 (\bar{r} - z)^2} + 
{(r - s)^2 \over (r - z)^2 (s - z)^2} - 
{(\bar{r} - s)^2 \over (\bar{r} - z)^2 (s - z)^2} \right] \, 
Q (z, \bar{r},\bar{s},s)\, S (r,z)  \nonumber \\
&+&
\left[{(r - \bar{r})^2 \over (r - z)^2 (\bar{r} - z)^2} + 
{(\bar{r} - \bar{s})^2 \over (\bar{r} - z)^2 (\bar{s} - z)^2} - 
{(r - \bar{s})^2 \over (r - z)^2 (\bar{s} - z)^2} \right] \,
Q (r, z,\bar{s},s)\, S (z,\bar{r})  
\nonumber \\
&+&
\left[ {(\bar{r} - \bar{s})^2 \over (\bar{r} - z)^2 (\bar{s} - z)^2} + 
{(s - \bar{s})^2 \over (s - z)^2 (\bar{s} - z)^2} -  
{(\bar{r} - s)^2 \over (s - z)^2 (\bar{r} - z)^2}  \right] \, 
Q (r, \bar{r},z,s)\, S (\bar{s},z)  \nonumber \\
&+&
\left[ {(r - s)^2 \over (r - z)^2 (s - z)^2} + 
{(s - \bar{s})^2 \over (s - z)^2 (\bar{s} - z)^2} -  
{(r - \bar{s})^2 \over (r - z)^2 (\bar{s} - z)^2} \right] \, 
Q (r, \bar{r},\bar{s},z)\, S (z,s)  \nonumber \\
&-&
\left[ {(r - \bar{r})^2 \over (r - z)^2 (\bar{r} - z)^2} + 
{(s - \bar{s})^2 \over (s - z)^2 (\bar{s} - z)^2} + 
{(r - s)^2 \over (r - z)^2 (s - z)^2} + 
{(\bar{r} - \bar{s})^2 \over (\bar{r} - z)^2 (\bar{s} - z)^2} \right] 
\, Q (r, \bar{r},\bar{s},s) \nonumber \\
&-&
\left[ {(r - s)^2 \over (r - z)^2 (s - z)^2} + 
{(\bar{r} - \bar{s})^2 \over (\bar{r} - z)^2 (\bar{s} - z)^2} -  
{(\bar{r} - s)^2 \over (\bar{r} - z)^2 (s - z)^2} - 
{(r - \bar{s})^2 \over (r - z)^2 (\bar{s} - z)^2}  \right]  \,
 S (r,s)\,  S (\bar{r},\bar{s}) \nonumber \\
&-&
\left[ {(r - \bar{r})^2 \over (r - z)^2 (\bar{r} - z)^2} + 
{(s - \bar{s})^2 \over (s - z)^2 (\bar{s} - z)^2} - 
{(r - \bar{s})^2 \over (r - z)^2 (\bar{s} - z)^2} -  
{(\bar{r} - s)^2 \over (\bar{r} - z)^2 (s - z)^2} \right] \, 
S (r,\bar{r})\,  S (\bar{s},s)
\!\!\Bigg> \!\!  \Bigg\}
\label{eq:Q_evo}
\eea
where the $S$ matrix is defined as 
\be
S (r, \bar{r}) \equiv {1\over N_c}  \Tr V_r \, V^\dagger_{\bar{r}} 
\label{eq:s}
\ee
We will refer to the first four lines in this equation as "real" and the 
last three terms as "virtual" terms in coordinate space. This is to distinguish 
them from the real and virtual terms in momentum space after we Fourier transform 
the equation since there is no one to one correspondence between the real and 
virtual terms in coordinate and momentum spaces. We have also verified that this 
equation is exact in the sense that there are
no $N_c$ suppressed terms in the equation itself (note that models used
to evaluate the color averaging denoted by $< \cdots >$ may introduce
sub-leading $N_c$ terms). It also agrees with the previous results for the
quadrupole evolution equation~\cite{jjmyk1,dmmx}. The $S$ matrix satisfies the BK evolution 
equation~\cite{bk} given by  

\be
{d\over dy} \left< S (r - s)\right>  = {N_c\, \alpha_s \over 2\pi^2} \;
\int d^2 z \,
 {(r - s)^2 \over (r - z)^2 (s - z)^2}\,\bigg[
\left< S (r -z)\right> \, \left< S (z - s)\right>  - 
\left<  S (r -s)\right> 
\bigg]
\label{eq:bk}
\ee
Unlike the dipole kernel in the BK equation which allows a probabilistic
interpretation in coordinate space, the same is not true in the quadrupole
evolution equation due to terms with negative signs. Even though the individual
kernels in eq. (\ref{eq:Q_evo}) are just the standard dipole kernels~\cite{dipole_model},  
it is still perhaps useful to explain in a more intuitive way, the various terms that 
appear in eq. (\ref{eq:Q_evo}). The first four lines in eq. (\ref{eq:Q_evo}) 
are the "real" corrections and come from the third and fourth terms in eq. (\ref{eq:eta}). 
One can rewrite any kernel in eq. (\ref{eq:Q_evo}) in a way which may look more familiar 
and facilitates the comparison with the standard dipole emission kernel.
For example, the kernel in the first line on the right hand side of eq. (\ref{eq:Q_evo}) 
can be written as as 
\be
\sim 2 \left[{1 \over (r - z)^2} - {(r - z)\cdot (\bar{r} -z) \over (r - z)^2 (\bar{r} - z)^2} 
- 
{(r - z)\cdot (s -z) \over (r - z)^2 (s - z)^2} 
+ 
{(\bar{r} - z)\cdot (s - z) \over (\bar{r} - z)^2 (s - z)^2} \right]
\label{eq:kernel_1}
\ee
with a similar form for all the other kernels. Here the first term corresponds 
to a gluon being radiated by a quark line represented by $V (r)$. If it is absorbed by
the same quark line in the amplitude, it leaves the quadrupole unchanged and will correspond
to a "virtual" correction. On the other hand if it is absorbed by the same quark line but
in the complex conjugate amplitude (so the gluon line crosses the cut), it will multiply
the quadrupole with the coordinate $r$ replaced by $z$ and a dipole with coordinates $r,z$.
This will be part of the "real" corrections. The second term above corresponds to the case 
when the quark line, represented by $V (r)$, in the quark anti-quark system represented by 
$V (r)$ and $V (\bar{r})$ radiates a gluon with transverse coordinate $z$. If the radiated
gluon does not cross the cut line and is absorbed by the anti-quark line at $\bar{r}$ it becomes
part of the "virtual" corrections. On the other hand if the radiated gluon at $z$ crosses the cut
and is then absorbed by an anti-quark line in the complex conjugate amplitude, it breaks the
original quadrupole into a quadrupole with coordinate $r$ replaced by $z$ and a dipole at $r,z$.
This is part of the "real corrections. All other terms have a similar interpretation.

To investigate the weak field limit of this evolution and to make our approximations more transparent, 
it is more useful to work with the $T$ matrices, defined as $T_Q \equiv 1 - Q$ and $T \equiv 1 - S$. 
It is easy to see that all kernels multiplying $1$ (when we switch from $Q,S$ to $T_Q,T$) add up to zero. 
Therefore, eq. (\ref{eq:Q_evo}) is re-written as 
\bea
&& {d\over dy} \left< T_Q (r, \bar{r},\bar{s},s)\right> = 
{N_c\, \alpha_s \over (2\pi)^2} \int d^2 z \Bigg\{ \Bigg<  \nonumber \\
&& \left[ {(r - \bar{r})^2 \over (r - z)^2 (\bar{r} - z)^2} + 
{(r - s)^2 \over (r - z)^2 (s - z)^2} - 
{(\bar{r} - s)^2 \over (\bar{r} - z)^2 (s - z)^2} \right] \!\! 
\bigg[T_Q (z, \bar{r},\bar{s},s) + T (r,z) -  T_Q (z, \bar{r},\bar{s},s)  T (r,z)\bigg] 
\nonumber \\
&+&
\left[{(r - \bar{r})^2 \over (r - z)^2 (\bar{r} - z)^2} + 
{(\bar{r} - \bar{s})^2 \over (\bar{r} - z)^2 (\bar{s} - z)^2} - 
{(r - \bar{s})^2 \over (r - z)^2 (\bar{s} - z)^2} \right] \!\! 
\bigg[T_Q (r, z,\bar{s},s) + T (z,\bar{r}) - T_Q (r, z,\bar{s},s)\, T (z,\bar{r})\bigg]  
\nonumber \\
&+&
\left[ {(\bar{r} - \bar{s})^2 \over (\bar{r} - z)^2 (\bar{s} - z)^2} + 
{(s - \bar{s})^2 \over (s - z)^2 (\bar{s} - z)^2} -  
{(s - \bar{r})^2 \over (s - z)^2 (\bar{r} - z)^2}  \right] \!\! 
\bigg[T_Q (r, \bar{r},z,s) + T (\bar{s},z) -  T_Q (r, \bar{r},z,s)\, T (\bar{s},z)\bigg]  
\nonumber \\
&+&
\left[ {(r - s)^2 \over (r - z)^2 (s - z)^2} + 
{(s - \bar{s})^2 \over (s - z)^2 (\bar{s} - z)^2} -  
{(r - \bar{s})^2 \over (r - z)^2 (\bar{s} - z)^2} \right] \!\! 
\bigg[T_Q (r, \bar{r},\bar{s},z) + T (z,s) - T_Q (r, \bar{r},\bar{s},z)\, T (z,s)\bigg]  
\nonumber \\
&-&
\left[ {(r - \bar{r})^2 \over (r - z)^2 (\bar{r} - z)^2} + 
{(s - \bar{s})^2 \over (s - z)^2 (\bar{s} - z)^2} + 
{(r - s)^2 \over (r - z)^2 (s - z)^2} + 
{(\bar{r} - \bar{s})^2 \over (\bar{r} - z)^2 (\bar{s} - z)^2} \right] 
\, T_Q (r, \bar{r},\bar{s},s) \nonumber \\
&-&
\left[ {(r - s)^2 \over (r - z)^2 (s - z)^2} + 
{(\bar{r} - \bar{s})^2 \over (\bar{r} - z)^2 (\bar{s} - z)^2} -  
{(\bar{r} - s)^2 \over (\bar{r} - z)^2 (s - z)^2} - 
{(r - \bar{s})^2 \over (r - z)^2 (\bar{s} - z)^2}  \right]  \!\! 
\bigg[T (r,s) +  T (\bar{r},\bar{s}) - T (r,s)\,  T (\bar{r},\bar{s})\bigg] 
\nonumber \\
&-&
\left[ {(r - \bar{r})^2 \over (r - z)^2 (\bar{r} - z)^2} + 
{(s - \bar{s})^2 \over (s - z)^2 (\bar{s} - z)^2} - 
{(r - \bar{s})^2 \over (r - z)^2 (\bar{s} - z)^2} -  
{(\bar{r} - s)^2 \over (\bar{r} - z)^2 (s - z)^2} \right] \!\! 
\bigg[T (r,\bar{r}) + T (\bar{s},s) - T (r,\bar{r})\,  T (\bar{s},s)\bigg]
\Bigg> \Bigg\} \nonumber\\
\label{eq:T_Q_evo}
\eea

\subsection{The weak field limits}

It is useful to consider the above equation for $T_Q$ in the weak field (dilute) 
limit where all sizes are much smaller than the inverse saturation scale, i.e., 
$|a - b| << {1\over Q_s}$ for any external coordinates $a,b$. In this limit the non-linear 
terms ($T_Q \, T$ and  $T\, T$) in eq. (\ref{eq:T_Q_evo}) may be dropped and we get 

\bea
{d\over dy} \left< T_Q (r, \bar{r},\bar{s},s)\right> \!\!&=&\!\! 
 {N_c\, \alpha_s \over (2\pi)^2} \!\!\int \!\! d^2 z \Bigg\{ \!\!\Bigg< \!\! 
%
\left[ {(r - \bar{r})^2 \over (r - z)^2 (\bar{r} - z)^2} + 
{(r - s)^2 \over (r - z)^2 (s - z)^2} - 
{(\bar{r} - s)^2 \over (\bar{r} - z)^2 (s - z)^2} \right] \!\! 
\bigg[T_Q (z, \bar{r},\bar{s},s) + T (r,z) \bigg] +  
\nonumber \\
&&
\left[{(r - \bar{r})^2 \over (r - z)^2 (\bar{r} - z)^2} + 
{(\bar{r} - \bar{s})^2 \over (\bar{r} - z)^2 (\bar{s} - z)^2}  - 
{(r - \bar{s})^2 \over (r - z)^2 (\bar{s} - z)^2} \right] \!\! 
\bigg[T_Q (r, z,\bar{s},s) + T (z,\bar{r}) \bigg]  + 
\nonumber \\
&&
\left[ {(\bar{r} - \bar{s})^2 \over (\bar{r} - z)^2 (\bar{s} - z)^2} + 
{(s - \bar{s})^2 \over (s - z)^2 (\bar{s} - z)^2} -  
{(s - \bar{r})^2 \over (s - z)^2 (\bar{r} - z)^2}  \right] \!\! 
\bigg[T_Q (r, \bar{r},z,s) + T (\bar{s},z) \bigg]  + 
\nonumber \\
&&
\left[ {(r - s)^2 \over (r - z)^2 (s - z)^2} + 
{(s - \bar{s})^2 \over (s - z)^2 (\bar{s} - z)^2} -  
{(r - \bar{s})^2 \over (r - z)^2 (\bar{s} - z)^2} \right] \!\! 
\bigg[T_Q (r, \bar{r},\bar{s},z) + T (z,s) \bigg]   - 
\nonumber \\
&&
\left[ {(r - \bar{r})^2 \over (r - z)^2 (\bar{r} - z)^2} + 
{(s - \bar{s})^2 \over (s - z)^2 (\bar{s} - z)^2} + 
{(r - s)^2 \over (r - z)^2 (s - z)^2} + 
{(\bar{r} - \bar{s})^2 \over (\bar{r} - z)^2 (\bar{s} - z)^2} \right] 
\, T_Q (r, \bar{r},\bar{s},s) - 
\nonumber \\
&&
\left[ {(r - s)^2 \over (r - z)^2 (s - z)^2} + 
{(\bar{r} - \bar{s})^2 \over (\bar{r} - z)^2 (\bar{s} - z)^2} -  
{(\bar{r} - s)^2 \over (\bar{r} - z)^2 (s - z)^2} - 
{(r - \bar{s})^2 \over (r - z)^2 (\bar{s} - z)^2}  \right]  \!\! 
\bigg[T (r,s) +  T (\bar{r},\bar{s}) \bigg]  - 
\nonumber \\
&&
\left[ {(r - \bar{r})^2 \over (r - z)^2 (\bar{r} - z)^2} + 
{(s - \bar{s})^2 \over (s - z)^2 (\bar{s} - z)^2} - 
{(r - \bar{s})^2 \over (r - z)^2 (\bar{s} - z)^2} -  
{(\bar{r} - s)^2 \over (\bar{r} - z)^2 (s - z)^2} \right] \!\! 
\bigg[T (r,\bar{r}) + T (\bar{s},s) \bigg]
\!\Bigg> \!\Bigg\}
\nonumber \\
\label{eq:T_Q_evo_lin}
\eea
To proceed further, we first consider the two-gluon exchange limit, i.e., the BFKL equation~\cite{bfkl}. 
Since $T_Q$ and $T$ include multiple gluon exchanges, we need
to linearize them, i.e., take the single (reggeized) gluon exchange limit. This corresponds
to expanding each of the Wilson lines in the definition of $T_Q$ and $T$ to first 
order in the gauge field $\alpha$ and then keeping terms of the order 
$\alpha^2$. In this limit (note the relative sign which appears when taking
both $\alpha$'s from either $V$'s or $V^\dagger$'s rather than taking one $\alpha$ from a $V$ 
and another $\alpha$ from a $V^\dagger$)
\be
T_Q (r,\bar{r},\bar{s},s) \rightarrow T (r,\bar{r}) + T (\bar{s},s) - T (r, \bar{s}) - T (\bar{r},s)
+ T (r,s) + T (\bar{r}, \bar{s})
\label{eq:T_Q_bfkl}
\ee
Using eq. (\ref{eq:T_Q_bfkl}) in both sides of eq. (\ref{eq:T_Q_evo_lin})
we get the BFKL equation for each $T$ of a given argument. For example, 
 \be
{d\over dy} \left< T (r,s)\right>  = {N_c\, \alpha_s \over 2\pi^2} \;
\int d^2 z \,
 {(r - s)^2 \over (r - z)^2 (s - z)^2}\,\bigg[
\left< T (r , z)\right>  + 
\left< T (z , s)\right> - 
\left< T (r , s)\right>   
\bigg]
\label{eq:T_evo}
\ee  
where $T$ in eq. (\ref{eq:T_evo}) and right hand side of (\ref{eq:T_Q_bfkl}) stands for 
\be
T (r, \bar{r}) \rightarrow  \Gamma (r - \bar{r}) \sim 
g^2\, \alpha^a (r) \alpha^a (\bar{r})
\label{eq:T_bfkl}
\ee
This limit was already considered in~\cite{dmmx} and the correspondence with BFKL was shown.
We also note that this relation still holds when the evolution equation is written in
terms of the color charge density $\rho$ rather than the gauge field $\alpha$.
  
The next interesting case is to consider $O (\alpha^4)$ and see whether 
our evolution equation reduces to the well-known BJKP equation governing the evolution of 
four reggeized-gluon state in the dilute limit. To do this, again we first ignore the 
non-linear terms in the evolution equation, then we expand the Wilson lines 
and keep terms of the order $\alpha^4$ in eq. (\ref{eq:T_Q_evo_lin}). 
Since the BJKP equation is written in momentum space, we will start by Fourier transforming
$T_Q$ (ignoring $T$ at the moment) to momentum space and disregard any contribution which leads to a vanishing external
momentum. We define~\footnote{We will use the notation $T_4$ here to denote the $ \sim O (\alpha^4)$ 
terms in the expansion of $T_Q$ so that $T_4 \equiv {1\over N_c}\,Tr\, [\alpha\,\alpha\,\alpha\,\alpha]$.}
\be
T_4 (l_1,l_2,l_3,l_4) \equiv \int d^2 r\, d^2 \bar{r}\, d^2 \bar{s}\, d^2 s\, 
e^{i (l_1 \cdot r \,+\, l_2 \cdot \bar{r} \,+\, l_3 \cdot \bar{s} \,+\, l_4 \cdot s)}\, 
T_4 (r,\bar{r},\bar{s},s) 
\ee
where $l_1,l_2,l_3,l_4$ are two-dimensional external transverse momenta satisfying overall
transverse momentum conservation so that there are only three independent momenta. This 
corresponds to having a choice in picking the origin of the coordinate space on the transverse
plane. One can then right away see that the last
term in (\ref{eq:kernel_1}) convoluted with $T_4 (z, \bar{r},\bar{s},s)$ will give a 
$\delta^2 (l_1) $ since it does not depend on coordinate $r$. A similar argument shows
that the last term in each kernel in the first $4$ lines in eq. (\ref{eq:T_Q_evo_lin}) (the
"real" terms) will lead to a delta function which sets one of the external momenta to zero. 
Since the external momenta of the reggeized-gluons are assumed to be finite (non-zero), all these terms
can be safely ignored. We now consider the contribution of the "virtual" terms, line $5$ 
in eq. (\ref{eq:T_Q_evo_lin}). Upon Fourier transforming, we get
\be
- 8\, {N_c\, \alpha_s \over (2\pi)^2} \int {d^2 p_t \over p_t^2}\,   T_4 (l_1,l_2,l_3,l_4) 
+ 4\, {N_c\, \alpha_s \over (2\pi)^2} \int {d^2 p_t \over p_t^2}\,   T_4 (p_t + l_1,l_2 - p_t,l_3,l_4) + \cdots
\label{eq:vir}
\ee
with a cyclic permutation of the external momenta in the second term understood. The first term is part
of the virtual corrections while the second term is part of the real corrections in momentum space. Let 
us consider now the contribution of "real" terms. Fourier transforming the non-zero terms in the first line 
of eq. (\ref{eq:T_Q_evo_lin}) gives 
\be
2 {N_c\, \alpha_s \over (2\pi)^2} \int d^2 p_t \left[
{p_t \cdot (p_t - l_1) \over p_t^2 (p_t - l_1)^2} T_4 (l_1,l_2,l_3,l_4) 
+ 2\, {p_t\cdot l_1 \over p_t^2 l_1^2} T_4 (p_t + l_1 , l_2 - p_t,l_3,l_4)  
\right]
\label{eq:line_1}
\ee
The first term in eq. (\ref{eq:line_1}) is part of the virtual corrections (in momentum space)
while the second term is part of the real corrections. With a slight rearrangement of the first term 
one can rewrite the contribution of the first line in  eq. (\ref{eq:T_Q_evo_lin}) as  
\be
2 {N_c\, \alpha_s \over (2\pi)^2} \int d^2 p_t \, \left\{
\left[{1 \over p_t^2} - {l_1^2 \over2\, p_t^2 (p_t - l_1)^2}\right]
\, T_4 (l_1,l_2,l_3,l_4) 
+ 2\, {p_t\cdot l_1 \over p_t^2 l_1^2} T_4 (p_t + l_1 ,l_2 - p_t,l_3,l_4)\right\}  
\label{eq:line_1_final}
\ee
It is clear that the first term in the square bracket in eq. (\ref{eq:line_1_final}) partially cancels the first term
in eq. (\ref{eq:vir}). This cancellation becomes complete when we include the similar contributions
from the lines $2-4$ in eq. (\ref{eq:T_Q_evo_lin}) so that the only virtual correction left so far is
the second term in the square bracket in (\ref{eq:line_1_final}). Including the contribution of
the second line to the real part (only the terms which lead to $T_4$ with the same argument, at the moment) 
gives
\bea
{d\over d y} T_4 (l_1,l_2,l_3,l_4) &=& {N_c\, \alpha_s \over \pi^2} \int d^2 p_t
\left[{1\over p_t^2} + {p_t\cdot l_1 \over p_t^2 l_1^2} - {p_t\cdot l_2 \over p_t^2 l_2^2}  
- {l_1\cdot l_2 \over l_1^2 l_2^2} \right]\, T_4 (p_t + l_1 ,l_2 - p_t,l_3,l_4) + \cdots \nonumber \\
&-&  
{N_c\, \alpha_s \over (2\pi)^2} \int d^2 p_t\, \left[{l_1^2 \over p_t^2 (l_1 - p_t)^2} + 
\{l_1 \rightarrow l_2, l_3, l_4\} \right] \,  T_4 (l_1,l_2,l_3,l_4)  
\label{eq:bjkp_alpha}
\eea
where $\cdots$ stands for real contributions obtained by appropriate permutation of the external momenta. 
Finally we note that the term proportional to $l_1\cdot l_2$ comes from keeping $ O(\sim \alpha^2)$ in
the expansion of $V_z$ and setting one of the other $V$'s to unity, for example, taking $V_{\bar{r}} =1$
and $\alpha^2 (z)$ in the first line of eq. (\ref{eq:T_Q_evo_lin}). It is clear that the virtual terms 
in eq. (\ref{eq:bjkp_alpha}) are already in exact agreement with one gets from BJKP 
equation~\cite{multigluon,chen-al} but the real terms look different. To show agreement of the real terms 
with the BJKP equation, we rewrite this equation for color charge density $\rho$ rather than
the gauge field $\alpha$ (this does not affect the virtual corrections). To this end, we note that the square 
bracket in the real term in eq. (\ref{eq:bjkp_alpha}) can be rewritten as
\be
\left[{1\over p_t^2} + {p_t\cdot l_1 \over p_t^2 l_1^2} - {p_t\cdot l_2 \over p_t^2 l_2^2}  
- {l_1\cdot l_2 \over l_1^2 l_2^2} \right] = {1\over 2} 
\left[ {(p_t + l_1)^2 \over p_t^2 l_1^2} + {(p_t - l_2)^2 \over p_t^2 l_2^2}  
- {(l_1 + l_2)^2 \over l_1^2 l_2^2}\right]\nonumber 
\ee
Recalling the relation between gauge field $\alpha$ and color charge density $\rho$,
\be
\alpha (p_t) \sim {\rho (p_t) \over p_t^2}
\label{eq:alpha_to_rho}
\ee
and defining $\hat{T}_4 (l_1,l_2,l_3,l_4) = {1\over N_c}\, Tr\, \rho (l_1) \rho (l_2) \rho (l_3) \rho (l_4)$, 
we multiply both sides of eq. (\ref{eq:bjkp_alpha}) with $l_1^2\, l_2^2\, l_3^2\, l_4^2$ which effectively
removes the external legs. Eq. (\ref{eq:bjkp_alpha}) can then be written as 
\bea
{d\over d y} \hat{T}_4 (l_1,l_2,l_3,l_4) &=& {N_c\, \alpha_s \over \pi^2} \int d^2 p_t
\left[{p^i\over p_t^2} - {(p^i - l_1^i) \over (p_t + l_1)^2}\right]\cdot
\left[{p^i\over p_t^2} - {(p^i - l_2^i) \over (p_t + l_2)^2}\right]\, 
\hat{T}_4 (p_t + l_1 ,l_2 - p_t,l_3,l_4) + \cdots \nonumber \\
&-&  
{N_c\, \alpha_s \over (2\pi)^2} \int d^2 p_t\, \left[{l_1^2 \over p_t^2 (l_1 - p_t)^2} + 
\{l_1 \rightarrow l_2, l_3, l_4\} \right] \,  \hat{T}_4 (l_1,l_2,l_3,l_4)  
\label{eq:bjkp_rho}
\eea   
This is our final result and corresponds to the evolution of $\hat{T}_4$ after one step in rapidity 
$y$ as depicted (the real part) in Fig. (1). We have checked that it agrees with the expressions 
given in~\cite{multigluon,chen-al}.

\begin{figure}[htb]
\includegraphics[width=12cm]{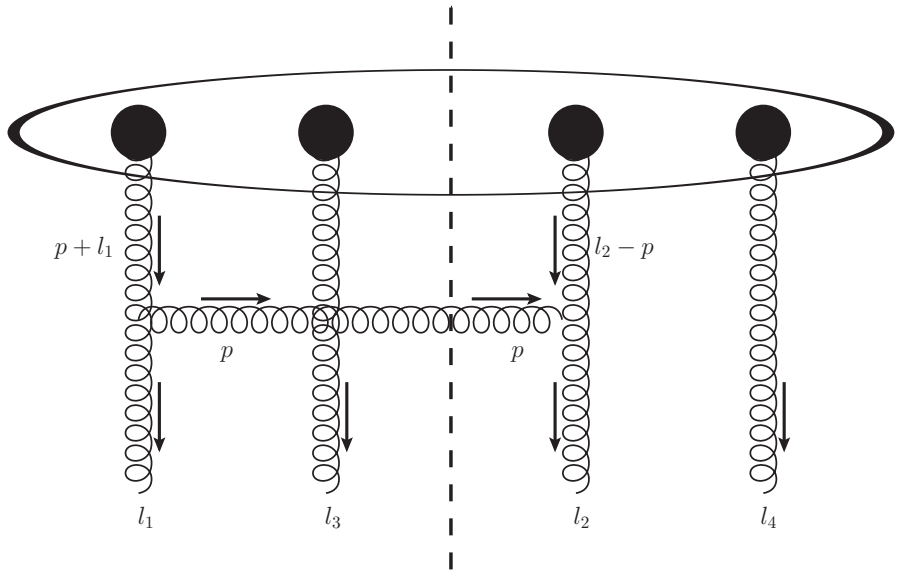}
\caption{Evolution of the four-point function $\hat{T}_4$ after one step in rapidity as given by eq. (\ref{eq:bjkp_rho}). 
Shown is one of the real diagrams only and the dashed line represents a cut.}
\label{fig:bjkp}
\end{figure}

There are several points that need to be clarified; first, we have completely disregarded the dipole
terms ($\sim T$) here even though they also contain $O (\alpha^4)$ terms. Since $T (r,s)$ depends only on 
two external transverse coordinates $r,s$,  $O (\alpha^4)$ terms will necessarily involve two pairs of 
gauge fields at the same point. Assuming rotational invariance on the transverse coordinate plane, 
this leads to setting two of the external momenta equal to each other which takes one back to the BFKL ladders.
Therefore, these terms are not relevant for our purpose. A second point is the color averaging denoted by $<\cdots >$. 
We have not made any assumptions about the color averaging~\cite{djp} and the evolution equation derived is independent of 
how one performs this averaging. Furthermore, the overall color structure of the equation seems to be more 
general than the BJKP equation since here one has a trace of four color matrices in the fundamental representation
on both sides of the equation. This trace could be written in terms of products of the group structure constants 
$\delta^{ab}, f^{abc}, d^{abc}$ whereas the BJKP equation is for the exchange of four reggeized-gluon state in 
a symmetric color singlet state. One expects that $\delta \, \delta$ terms would lead to a topology which
is equivalent to exchange of two independent BFKL pomerons which would then be disregarded. Therefore, one would only 
consider the color symmetric structures involving $d$'s.  

In summary, we have shown in this preliminary study that the JIMWLK evolution equation for the quadrupole operator can be 
reduced to the BJKP equation for the real part of the four reggeized-gluon exchange amplitude. To do this, we first 
ignore the non-linear (recombination) terms in quadrupole evolution equation, and then expand the Wilson lines in terms 
of the gauge field (or equivalently, the color charge density). This approximation should be valid when the external momenta
are larger that the saturation scale, i.e., in the dilute region. The quadrupole evolution equation reduces to a sum of
independent BFKL equations in $O (\rho^2)$ and to the BJKP equation when one looks at the terms of order  $\sim \rho^4$. 
This suggests that the JIMWLK evolution equation for the $n$-pole operator 
${1\over N_c} \, < Tr\, V (x_1)\, V^\dagger (x_2) \cdots  V^\dagger (x_n) >$ 
in the linear limit (dilute region) may be equivalent to the BJKP heirarchy for the imaginary part of the $n$ reggeized-gluon exchange 
amplitude. This would be very useful since there is much that is known about the BJKP equation and its properties but not 
much is known about the properties of the JIMWLK equation in analytic form. Proving the equivalence between linearized 
JIMWLK and BJKP equations may not be so difficult since the JIMWLK evolution equation for  
${1\over N_c} \, < Tr\, V (x_1)\, V^\dagger (x_2) \cdots  V^\dagger (x_n) >$  can almost be written down by inspection in
analogy with the pattern seen in eq. (\ref{eq:Q_evo}). The problem reduces to keeping track of which quark line radiates a 
gluon and counting all the possibilities since all emission kernels are just the standard dipole kernel. It would also 
be interesting to investigate the connection between the non-linear terms in the JIMWLK equation and multi-pomeron 
vertices employed in reggeized-gluon approach to high energy scattering. These issues are beyond the scope of this 
preliminary work and will be reported elsewhere.

\begin{acknowledgments}
We thank F. Dominguez, A. Dumitru, Y. Kovchegov, A. Mueller and B. Xiao for useful 
discussions. This work is supported by the DOE Office of Nuclear Physics
through Grant No.\ DE-FG02-09ER41620 and by The City University of
New York through the PSC-CUNY Research Program, grant 62625-41. Figures are made
using JaxoDraw~\cite{jaxodraw}.
\end{acknowledgments}

\end{document}